\documentclass[aps,preprint]{revtex4}
\usepackage{graphicx,amsmath,amssymb}

\newcommand{\bea}{\begin{eqnarray}}
\newcommand{\eea}{\end{eqnarray}}
\newcommand{\be}{\begin{equation}}
\newcommand{\ee}{\end{equation}}

\begin{document}

\begin{titlepage}
\title{Fragility index of a simple liquid from structural inputs}
\author{Leishangthem Premkumar and Shankar P. Das}
\affiliation{School of Physical Sciences,\\
Jawaharlal Nehru University,\\
New Delhi 110067, India.\\}

\setcounter{equation}{0}

\begin{abstract}

We make a first principle calculation of the fragility index $m$ of
a simple liquid using the structure of the supercooled liquid as an
input. Using the density functional theory (DFT) of classical
liquids, the configurational entropy ${\cal S}_c$ is obtained for
low degree of supercooling. We extrapolate this data to estimate the
Kauzmann temperature $T_\mathrm{K}$ for the liquid.  Using the
Adam-Gibbs relation, we link the configurational entropy ${\cal
S}_c$ to the relaxation time. The relaxation times are obtained from
direct solutions of the equations of fluctuating nonlinear
hydrodynamics (FNH). These equations also form the basis of the mode
coupling theory (MCT) for glassy dynamics. The fragility index for
the supercooled liquid is estimated from analysis of the curves on
the Angell plot.

\end{abstract}

\vspace*{.8cm}

\pacs{05.10.}

\maketitle
\end{titlepage}

\section{Introduction}

The thermodynamic equilibrium state of a liquid is characterized in
terms of a few characteristic  variables like temperature $T$,
pressure $P$, volume $V$, equilibrium density $\rho_0$. The
equilibrium liquid state is isotropic in a time averaged sense and
the constituent particles have random motion. The isotropic liquid
transforms in to a crystalline solid when its temperature $T$ falls
below a characteristic value $T_m$, termed as the freezing point of
the liquid at the corresponding pressure $P$. The isotropic symmetry
of the normal liquid state is spontaneously broken at $T_m$. The
crystalline state has characteristic long range order. The
transformation of the liquid in to crystalline solid involves
absorption of latent heat. Freezing process is distinct from the
condensation of the gaseous state in to the liquid state. The
density functional theory (DFT) presents an order parameter theory
\cite{RY} for freezing, using the equilibrium density as the
relevant variable. In DFT, the thermodynamic properties of the
inhomogeneous crystalline state are obtained in terms of the
corresponding properties of the homogeneous liquid state. The
thermodynamics of the dense uniform liquid is well understood in a
microscopic approach through integral equations theories
\cite{hansen} or simulations. The interaction potential between the
liquid particles constitute the microscopic level description of the
many particle system. The basic characteristics of the two body
potential for which a crystalline state appears (under appropriate
conditions of density and temperature) include (a) a strongly
repulsive part at short range and an (b) an attractive part
effective over long range. The Hamiltonian for the many particle
system is written in a harmonic expansion around the equilibrium
sites which correspond to the minimum potential energy
configuration. The attractive part of the potential seemingly
appears to play an important role in stabilizing the solid in a
crystalline state in which the individual particles localized around
their mean positions.

In the classical DFT, the free energy of an inhomogeneous system is
obtained as a functional of the one particle density $\rho({\bf
x})$\cite{evans_1979}. The density function is expressed in terms of
a suitable set of parameters which are treated as the order
parameters of the freezing transition. The free energy functional is
minimized with respect to these parameters. A very successful
prescription of density distribution for the crystalline case is
obtained from the superposition of Gaussian density
profiles\cite{PT} centered on a lattice with long range order,
\begin{equation}
\label{dens} \rho (\vec r)=\sum_i \phi_0( | \vec r-\vec R_i | )
\end{equation}
where the $\{ {\vec R_i} \} $ denotes the underlying lattice and the
function $\phi_0$ is taken as the isotropic Gaussian $ \phi_0(r)=
(\frac{\alpha}{\pi})^\frac{3}{2}e^{-\alpha r^2} $. The thermodynamic
properties of the system are computed assuming the latter to be in a
single phase, {\em i.e.}, either liquid or crystal. The density
functional approach is mean field like since it ignores the effects
of fluctuations. At  a given density by locating the free energy
minimum the corresponding structure is identified as the stable
thermodynamic state. For high temperatures, the homogeneous liquid
state is more stable while at low temperatures the crystalline state
with long range order is more stable. For the simple Lennard-Jones
system that we consider here, the face centered cubic (fcc)
structure is more stable.


The present paper focuses on the statistical mechanics of liquids
below freezing. Almost all liquids can be supercooled with varying
degrees of ease, below the freezing point $T_m$ without transforming
in to an ordered crystalline state. The liquid continues to remain
in the amorphous state and its characteristic relaxation time
drastically increases with lowering of temperature. The so called
glass transition point $T_\mathrm{g}$ which denotes the
vitrification process, is defined as the temperature at which the
relaxation time of the under-cooled liquid reaches the laboratory
time scales. The supercooled liquid at this stage behaves like a
solid with elastic properties. Unlike freezing process, this
transformation is not associated with any latent heat absorption and
there is a drop in the specific heat due to absence of the
translational degree of freedom. The free energy of the supercooled
liquid is expected not to show any discontinuous change through the
glass transition. At deep supercooling, liquid can remain trapped in
a metastable state having free energy intermediate between the
liquid and the crystalline state. Since there are a large number of
available metastable structures in which the under-cooled liquid can
be trapped, a considerable entropic drive is present for the
process. There have been theories for the vitrification process
built on possible scenarios of first order transitions with the
special situation of a large number of available metastable states
\cite{mosaic}.

An instructive  plot of the data of glassy relaxation was made by
Angell \cite{angell} of relaxation time $\tau_\alpha$  vs. inverse
temperature $T_\mathrm{g}/T$ scaled with $T_\mathrm{g}$ on a
logarithmic scale. The nature of increase of relaxation time with
fall of temperature in glassy systems is non-universal. One extreme
is a slow growth of $\tau_\alpha$  with lowering of temperature $T$
over the temperature range $T>T_\mathrm{g}$ followed by very sharp
increase within a small temperature range close to $T_\mathrm{g}$. A
more uniform increase is seen over the whole temperature range for
strong liquids like ${\rm B}_2{\rm O}_3$ or Si${\rm O}_2$. This
behavior has been quantified by defining a fragility parameter $m$
as the slope of the viscosity-temperature curve as \cite{bohmer1}.
\be \label{frag-defn}
m=\frac{d\log_{10}{\tau_\alpha}}{d(T_\mathrm{g}/T)}{\Big
|}_{T=T_\mathrm{g}}~~. \ee
Thus for example, $o$-terphenyl and Si${\rm O}_2$ denote two extreme
cases of fragile and strong systems with $m$ values 81 and 20
respectively. At the extreme fragile end the change of relaxation
time is extremely dramatic growing by many orders of magnitude
within a very narrow temperature range.


The supercooled liquid acts like a frozen solid over time scales of
structural relaxation and have only vibrational motion around a
frozen structure \cite{bpeak-pre,bpeak-pla}. The difference of the
entropy of the supercooled liquid from that of the solid having only
vibrational motion represents the entropy due to large scale motion
of the particles and is identified as the configurational entropy
${\cal S}_c$ of the liquid. A rapid disappearance of the
configurational entropy of the disordered liquid occurs on
approaching the glass transition point. This so called ``entropy
crisis" poses an important question essential for our understanding
the physics of glass transition and the divergence of relaxation
time at $T_\mathrm{g}$. Apart from having characteristic large
viscosity, the supercooled liquid shows a discontinuity in specific
heat $c_p$ at $T_\mathrm{g}$ due to freezing of the translational
degrees of freedom in the liquid. The above described features are
almost universally observed in all liquids. The Kauzmann temperature
$T_\mathrm{K}$ is the temperature at which the extrapolated value of
${\cal S}_c$ goes to zero and marks a possible limiting temperature
for the existence of the supercooled liquid phase. Below
$T_\mathrm{K}$  we have the paradoxical situation in which the
entropy of the disordered state becomes less than that of the
crystal. The original hypothesis due to Kauzmann proposes eventual
crystallization in the supercooled liquid at very low temperatures
as a  possible way out. Another possible explanation of the Kauzmann
paradox could be that the simple extrapolation of high temperature
result to very low temperature is not correct and the entropy
difference between supercooled liquid and crystal remains finite
till very low temperature \cite{Langer_2007,Donev_2006}, finally
going to zero only near $T=0$.


The theory of the supercooled liquid primary deals with two broad
aspects of the metastable state. These respectively refer to the
thermodynamics property like  the configurational entropy and the
slow dynamics characteristic of the glassy state. With increased
supercooling the relaxation time for the liquid sharply increases.
Relaxation in the present context is meant for a typical fluctuation
around the disordered liquid state at a temperature $T<T_m$.
Dynamics of the deeply supercooled liquid changes over from a
continuous motion of its particles to transport by activated hopping
over barriers that develops at low temperature. In structural glass
this occurs even without any quenched impurities, {\em i.e.}, the
slow dynamics is self generated. In the Adam-Gibbs theory
\cite{agibbs} the growth of the relaxation is linked to the
configurational entropy ${\cal S}_c$ of the supercooled liquid. The
idea that energy barriers build up to resist molecular rearrangement
in the jammed state has been used in the Adam-Gibbs theory to
understand the development of very long relaxation times in the
deeply supercooled state \cite{dim-cdg}. Since relaxation in the
system occurs through thermally assisted hopping over the barrier
(=$E_B$ say), the probability of such a jump will be controlled by
the Boltzmann factor $\exp(-E_B/k_BT)$. Thus estimation of the
relaxation time is closely linked to that of the energy barrier
$E_B$ which must be overcome so that a local fluctuation can relax.
Using these ideas it is argued \cite{agibbs} that the relaxation
time $\tau$ is linked to the configurational entropy ${\cal S}_c$ at
temperature $T$ of the liquid through the relation
\be \label{ag-reln} \tau=\tau_0 \exp \left [ \frac{A_0}{T{\cal S}_c}
\right ] ~~,\ee
where $A_0$ is a constant. As $T{\rightarrow}T_\mathrm{K}$, the
configurational entropy ${\cal S}_c\rightarrow{0}$ and hence
$\tau{\rightarrow}\infty$. Thus assuming a linear temperature
dependence of $T{\cal S}_c$ near $T_\mathrm{K}$ we can identify
$T_\mathrm{K}$ with the temperature $T_0$ of the standard
Vogel-Fulcher dependence of relaxation $\tau=\tau_0\exp(A/(T-T_0)$
\cite{vogel}. This equality between $T_0$ and $T_\mathrm{K}$
suggests a link of deeper significance on considering the fact that
the physics of the two temperatures are very different. $T_0$
represents the temperature at which the relaxation time for the
supercooled liquid diverges and basically relates to the dynamics.
On the other hand the Kauzamann temperature $T_\mathrm{K}$ is
related to the vanishing of the thermodynamic property of
configurational  entropy of the metastable liquid. Linking of the
sharp in crease of relaxation time to the entropy crisis signifies
effects of structure on the dynamics \cite{entr-jcp}.


For studying the classical liquid at densities beyond freezing
point, the DFT \cite{ys-pr,lowen-pr,mybook-cam} and the MCT
\cite{my_rmp,reichman-chr} has been two primary tools. In the
present paper we use of both approaches to study the properties of
the supercooled liquid. The density functional methods has been
adopted for studying the thermodynamic properties of the liquid
while mode coupling theories offers a microscopic model for slow
dynamics of the metastable liquid approaching glass transition. In
its simplest form the theory predicts a sharp dynamic transition
around a temperature $T_c$ higher than $T_\mathrm{g}$. This is a
transition from the ergodic liquid state to a nonergodic state in
which long time limits of the density correlation function does not
decay to zero. Around $T_c$, scaling behavior the dynamics of the
liquid undergoes a qualitative change. Using only structural inputs,
scaling of the non ergodicity parameter \cite{cusp_jcp} and growing
dynamic length scale \cite{dlength,biroli-epl} have been studied
around the MCT transition point.

Similar to the studies of the freezing transition, the DFT methods
have been applied to model the supercooled liquid below the freezing
point $T_m$, having aperiodic structures \cite{ysingh,
dasgupta_1992,PRL_DFT}. The inhomogeneous states are characterized
by localized density profiles (over suitable time scales) around a
disordered set of lattice points. For the aperiodic structure, the
corresponding $\{{\bf R}_i\}$ in the definition (\ref{dens}) for the
density constitute a random structure. The quantity $\alpha$ in Eq.
(\ref{dens}) is the variational parameter \cite{chandan-sood}.
Inverse of $\alpha$ characterizes the width of the peak and
therefore signifies the degree of mass localization in the system.
The homogeneous liquid state is characterized by the limit
$\alpha{\rightarrow}0$ and each Gaussian profile provides the same
contribution in the sum at all spatial positions. The metastable
states are identified as minima of the free energy, intermediate
between a crystal and a homogeneous liquid state.

Next we consider the model for understanding the dynamics, {\em
i.e.}, the mode coupling models. Recently the relaxation time of a
simple liquid has been calculated \cite{pre_barrat} from a direct
solution of the equations of fluctuating nonlinear hydrodynamics
(FNH). These equations are also the starting point of mode coupling
theory. On the other hand using the density functional methods we
compute here the configurational entropy ${\cal S}_c$ in the region
close to $T_m$. In the present paper we use the temperature
dependence of the relaxation time and the Adam-Gibbs relation
involving the configurational entropy ${\cal S}_c$ (in the higher
temperature range) to estimate the fragility index of a simple
Lennard-Jones liquid. ${\cal S}_c$ is estimated using the structure
of the liquid. Hence the present calculation only requires as an
input the basic interaction potential in terms of which the
structure factors are obtained. The paper is organized as follows.
In Sec. II, we present the respective models studied for the
thermodynamics and the dynamics separately. In Sec. III the
numerical results obtained for the configurational entropy is
checked for the validity of the Adam-Gibbs relation with the use of
the relaxation time. We then explore the Angell plots of the model
studied. The paper ends with a discussion section.

\section{Model studied}

We are dealing in this paper with two types of microscopic models
for the description of the metastable liquid. First, the density
functional model which provides a description of the properties
related to thermodynamics. Second, the equations of fluctuating
nonlinear hydrodynamics for the supercooled liquid signifying the
underlying conservation laws for the many particle system. We
briefly describe these two approaches in this section.

\subsection{Model for thermodynamics}

First we briefly outline the construction of the proper free energy
functional $F[\rho]$ corresponding to the ensemble in which the
average density is computed. This functional is then used to
determine the appropriate parameters for the inhomogeneous density
function $\rho({\bf x})$ at equilibrium. This is done by satisfying
the extremum principle for $F[\rho]$. For the canonical ensemble it
is the Helmoholtz free energy functional $F[\rho]$ which is to be
minimized to identify the equilibrium state. The free energy of the
liquid is obtained as a sum of two parts - the ideal gas term and
the interaction term,
\begin{equation}
\label{totf_ide_int} F[\rho]=F_\mathrm{id}[\rho] +
F_\mathrm{ex}[\rho] .
\end{equation}
The ideal gas part of the free energy $F_{id}$ for the non-uniform
density is obtained as
\begin{equation}\label{fideg}
F_{id} [ \rho({\bf r})] = k_B T\int d{\bf r}  \rho({\bf
r})\left(\ln[\wedge^{3} \rho({\bf r})]-1\right).
\end{equation}
$\wedge$ represents the thermal wavelength appearing due to the
momentum variable integration in the partition function. The RHS of
Eq. (\ref{fideg}) is a simple generalization of the ideal gas part
of the free energy for the nonuniform density, {\em i.e.,} $\rho
\rightarrow \rho (x)$. The interaction part is evaluated using the
standard expression for the Ramakrishnan-Yussouff (RY) functional
\cite{RY} involving a functional Taylor series expansion in terms of
the density fluctuation $\delta \rho(\vec r) = \rho(\vec r) -\rho_o
$ around liquid phase of average density $ \rho_o $,
\bea \label{Fex-RY} F_\mathrm{ex} &=& F_\mathrm{ex}(\rho_0) -\int
d{\bf x}_1 c^{(1)}({\bf x}_1;\rho_0) \delta \rho({\bf x}) \\
&-& \frac{1}{2}\int d {\bf x}_1 \int d {\bf x}_2 c^{(2)} ({\bf
x}_1,{\bf x}_2 ;\rho_0) \delta \rho({\bf x}_1) \delta  \rho({\bf
x}_2) + .. \nonumber  \eea
The series involve the functions $c^{(i)}$'s defined in
(\ref{dirc-c-defn}) at the liquid state density $\rho({\bf
x})=\rho_0$. For a uniform homogeneous liquid, $\rho_0$ will be
independent of position. We use the following definitions for the
direct correlation functions $c^{(i)}$s as the successive functional
derivatives of $F_\mathrm{ex}$ evaluated at the liquid state density
$\rho_0$,

\begin{equation}
\label{dirc-c-defn} c^{(i)}({\bf x}_1, ...,{\bf x}_i;\rho_0)= -\left
[\frac{\delta^i F_\mathrm{ex}}{\delta \rho({\bf x}_1)...\delta
\rho({\bf x}_i)} \right ]_{\rho=\rho_0}
\end{equation}
For practical calculations one usually adopts the simplest
approximation keeping only up to the second order term ($i=2$) in
the expansion for the direct correlation function. The functional
extremum principle now reduce to the form for the canonical ensemble
as
\begin{equation}
\label{min-con3} \ln[\wedge^3\rho({\bf x})]- c^{(1)}({\bf x};
\rho({\bf x}))+\beta\phi=0~~,
\end{equation}
where $\phi({\bf x})$ is the external potential. Using the result
(\ref{min-con3}) we obtain for the equilibrium density $\rho({\bf
x})$
\begin{equation}
\label{dens-genrl} \rho({\bf x}) = z\exp[-\beta\phi({\bf x})+
 c^{(1)}({\bf x}; \rho_0 ({\bf x}))]~~,
\end{equation}
where $z=\wedge^{-3}$ in this case. The quantity $c^{(1)}({\bf x};
\rho({\bf x}))$ acts as a one body potential due to the interaction
between the fluid particles. The higher order direct correlation
functions are defined in terms of functional derivatives of
$c^{(1)}$ with respect to $\rho({\bf x})$. Making a simple Taylor
expansion for $c^{(1)}({\bf x}; \rho({\bf x}))$ around its value
$c_l$ in the uniform liquid state of density $\rho_0$, we obtain,
\bea \label{c_l-exp} c^{(1)}[{\bf x}_1; \rho({\bf x}_1] &=&
c_l(\rho_0) + \int d {\bf x}_2 c^{(2)} [{\bf x}_1,{\bf x}_2 ;\rho_0]
\delta
\rho({\bf x}_2)  \\
&+&  \frac{1}{2} \int d {\bf x}_2 d{\bf x}_3 c^{(3)} [{\bf x}_1,{\bf
x}_2,{\bf x}_3 ;\rho_0] \delta \rho({\bf x}_2)\delta \rho({\bf
x}_3)+....\nonumber \eea
where $\delta \rho({\bf x})= \rho({\bf x}) - \rho_0$ is the
fluctuation of the equilibrium density in the inhomogeneous solid
state from that of the liquid state. The two point function function
$c^{(2)}({\bf x}_1, {\bf x}_2)$ is related to the pair correlation
function $g^{(2)}({\bf x}_1, {\bf x}_2)$ in the fluid by a relation
which reduces to the Ornstein-Zernike relation for the uniform
liquid. For the uniform liquid in absence of any external field we
have from Eq. (\ref{dens-genrl}) for the uniform density
$\rho_0=z\exp(c_l)$. The inhomogeneous density function $\rho({\bf
x})$ is then obtained in terms of the corresponding one particle
direct correlation function $c^{(1)}(r)$,
\be \label{DFT_eqn} \rho({\bf x}) = \rho_0 \exp \left \{
c^{(1)}({\bf x}; \rho({\bf x}))-c_l-\beta\phi({\bf x}) \right \} ~~.
\ee
The equilibrium density is therefore obtained as
\be \label{dens-int}\rho({\bf x}_1) = \bar{\rho}_0({\bf x}_1) \exp
\left [ \int d {\bf x}_2 c^{(2)} ({\bf x}_1,{\bf x}_2 ;\rho_0)
 \delta  \rho_0({\bf x}_2) \right ]
\ee
where  we identify $\bar{\rho}_0({\bf x}) = z\exp[-\beta\phi({\bf
x})+ c_l({\bf x})]\equiv \rho_0\exp[-\beta\phi({\bf x})]$. For Eq.
(\ref{dens-int}) the trivial solution is then the uniform density
$\rho({\bf x}_1) = \rho_0$ in absence of any external field $\phi$.
The solution of Eq. (\ref{dens-int}) is the starting point for the
subsequent analysis for testing the possibility of an inhomogeneous
density state. The two point kernel function $c^{(2)} ({\bf
x}_1,{\bf x}_2 ;\rho_0)$ which is defined in terms of the functional
derivative of the one body potential $c^{(1)}$ is required to
completely specify the equation (\ref{dens-int}) for the
inhomogeneous density.

\subsection{Appropriate free energy functional}

The density functional which is minimized with respect to the
density functions $\rho({\bf x})$ is obtained here for the constant
$NVT$ ensemble. Both the homogeneous and the inhomogeneous states
are at the same temperature and volume and number of particles. The
corresponding thermodynamic potential which is minimized is the
Helmholtz free energy. The difference between the free energy
functionals in the inhomogeneous state with density $\rho({\bf x})$
and the homogeneous liquid state with density $\rho_0$ ( in absence
of the external potential $\phi({\bf x})$) is obtained as,
\begin{eqnarray}
\label{gpot-dif} \Delta{\cal F} &\equiv& {\cal F}[\rho({\bf x})] -
{\cal F}[\rho_0] = \Delta F_\mathrm{id}[\rho({\bf x})] +
\Delta F_\mathrm{ex}[\rho({\bf x})] \nonumber \\
&+&\int d{\bf x}_1(\rho(x_1)-\rho_0)\phi(x_1).
\end{eqnarray}
The difference $\Delta F_\mathrm{id} = F_\mathrm{id}[\rho({\bf x})]
- F_\mathrm{id}(\rho_0)$ in the ideal gas part of the free energy is
directly calculated from (\ref{f_ide}). The difference $\Delta
F_\mathrm{ex} =F_\mathrm{ex}[\rho({\bf x})]- F_\mathrm{ex} (\rho_0)$
between the excess free energies of the liquid and solid states is
expressed as a functional Taylor expansion in the density
fluctuations $\delta \rho({\bf x})=\rho({\bf x})-\rho_0$ from
(\ref{Fex-RY}). Using these results, we obtain the free energy
difference between the crystalline and liquid state as,
\begin{eqnarray}
\label{gpot1-dif} \Delta{\cal F} &=& \int d{\bf x}_1 \Bigg [
\rho({\bf x}_1)\ln\left [\frac{\rho({\bf x}_1)}{\rho_0}\right]-
{\delta}\rho({\bf x}_1)\left \{ 1+\ln (\rho_0\wedge^3)-c^{(1)}({\bf
x}_1;\rho_0)
+\phi(x_1) \right \}\Bigg] \nonumber \\
&-& \frac{1}{2}\int d {\bf x}_1 \int d {\bf x}_2 c^{(2)} ({\bf
x}_1,{\bf x}_2 ;\rho_0) \delta  \rho({\bf x}_1)
\delta  \rho({\bf x}_2) -...\nonumber \\
&=& \int d{\bf x}_1 \Bigg [ \rho({\bf x}_1)\ln\left [\frac{\rho({\bf
x}_1)}{\rho_0}\right]- \frac{1}{2}\int d{\bf x}_1  \int d {\bf x}_2
c^{(2)} ({\bf x}_1,{\bf x}_2 ;\rho_0) {\delta}\rho({\bf
x}_1){\delta}\rho({\bf x}_2) \Bigg ]~~~.
\end{eqnarray}
In reaching the above equation we have applied the extremum
condition (\ref{min-con3})  for the liquid state {\em i.e.},
\begin{equation}
\ln \rho ({\bf x}_1)-c^{(1)}({\bf x};\rho_0) +\phi(x)=0
\end{equation}
as well as the fact that for the canonical ensemble the total number
of particles are constant.

The procedure followed to compute the free energy for the
supercooled liquid would be to first identify the minimum of the
$\Delta {\cal F}$ with respect to $\alpha$ since the
$\alpha{\rightarrow}0$, the value of ${\cal F}$ is the liquid state
free energy. Once the optimum $\alpha$ is identified the
corresponding value of ${\cal F}$ for the optimum density gives the
free energy for the inhomogeneous state to leading order in density
fluctuations.

\subsection{Model for the Dynamics}

The slow dynamics of a dense liquid is generally studied in terms of
the correlation of density fluctuations which occur in the strongly
interacting many particle system. The structural relaxation is best
understood in terms of the two point dynamic correlation function
${C}(q,t_1,t_2)$ of density fluctuations at times $t_1$ and $t_2$,
corresponding to wave vector $q$. The correlation function is
defined in the normalized form
\begin{equation} \label{dacf-def}
{C}(q,t_1,t_2)= \frac{<\delta \rho(q,t_1) \delta
\rho(-q,t_2)>}{<\delta \rho(q,t_2) \delta \rho(-q,t_2)>}~~.
\end{equation}
For the equilibrium state, time translational invariance holds and
$C(t_1,t_2)$ is a function of $(t_1-t_2)$ only. The long time limit
of the time correlation of density fluctuations is treated as an
order parameter in the mode coupling theory (MCT) of glassy
dynamics. This quantity, termed as the nonergodicity parameter
(NEP), makes a discontinuous jump from being zero in the liquid
state to a nonzero positive value at the ergodic-nonergodic (ENE)
transition of MCT. The corresponding temperature $T_c$ identified
with the sharp transition signifies a point at which a qualitative
change occurs in the moderately supercooled regime. $T_c$ lies in a
temperature range between the freezing temperature $T_m$ and the
glass transition temperature $T_{\mathrm g}$. The sharp ENE
transition is smoothed off in a complete analysis of the
nonlinearities in the equations which control the dynamics of
density fluctuations. However the qualitative change in the dynamics
in the initial stages of supercooling, around $T_c$ are described in
terms of the basic equations of FNH. The model equations of MCT also
follows from these equation which are plausible generalizations of
equations of hydrodynamics extended to small wave lengths. These
equations have been solved numerically\cite{pre_barrat,cdg_valls}
and the relaxation times obtained are in good agreement with
simulations results on similar systems.

For an isotropic liquid, the model equations of FNH for the mass
density $\rho$ and momentum density \textbf{g}\cite{DM} in  the
simplest form are as follows :
\bea \label{cont} &&\frac{\partial\rho}{\partial{t}} + {\bf
\nabla}.{\bf
g} = 0, \\
\label{g_eq} && \frac{\partial g_{i}}{\partial t} + \rho \nabla_{i}
f(r,t) - L_{ij} \frac{g_j}{\rho} = \theta_{i}~~. \eea
The correlations of the Gaussian noise $\theta_{i}$ are related to
the bare damping matrix $L^0_{ij}$ \cite{boon-yip},
\be \label{noise_cor} \left\langle \theta_{i}(x,t)\theta_j
({x^\prime}t^\prime)\right\rangle = 2k_{B}TL^0_{ij}
\delta(t-t^\prime)\delta(x-x^\prime). \ee
For an isotropic liquid, the bare transport coefficients are
obtained as,
\be \label{baretr} L^0_{ij} = (\zeta_{0} +
\eta_{0}/3)\delta_{ij}\nabla^{2} + \eta_{0}\nabla_{i}\nabla_{j} \ee
where  $\zeta_{0}$ and $\eta_0$ respectively denote the bare bulk
and shear viscosities. For the glassy dynamics we focus on the
coupling of slowly decaying density fluctuations present in the
pressure functional, represented by the second term on the LHS of
Eq. (\ref{g_eq}). The nonlinear contribution in this term is
obtained with the function $f(r,t)$. The latter is  presented as a
convolution
\be \label{convol} f(\textbf{r},t) = m^{-1}\int d\textbf{r}
c(\textbf{r}-\textbf{r}^{'}) \delta \rho(\textbf{r}^{'},t). \ee
If we replace $\rho$ by $\rho_0$ in the RHS of Eq. (\ref{g_eq}) then
we have a dynamics linearized in density fluctuations. The above
described FNH equations are solved numerically on a grid. The direct
correlation function $c(r)$ is used as an input for solving the FNH
equations and the noise averaged correlation function $C(q,t)$ of
density fluctuations are obtained \cite{pre_barrat}. With the
thermodynamic property, {\em i.e.}, the free energy ${\cal F}$ being
known using the classical DFT methods outlined above and the
dynamics properties {\em i.e.}, relaxation time obtained from the
solutions of the equations of FNH, the Adam-Gibbs relation can be
tested near the temperature $T_c$.

\section{Numerical results}

In DFT, the free energy is expressed as a functional of the density
$\rho({\bf x})$ which incorporates two key properties of the solid
state. First, the extent of mass localization in the system is
denoted by the width parameter $\alpha$ defined in the Eq.
(\ref{dens}). Second, the underlying lattice $\{{\bf R}_i\}$ on
which the Gaussian density profiles are to be centered. Both of
these properties are treated as control parameters of DFT.

\begin{figure}[htbp!]
\centering
\includegraphics[width=0.6\textwidth]{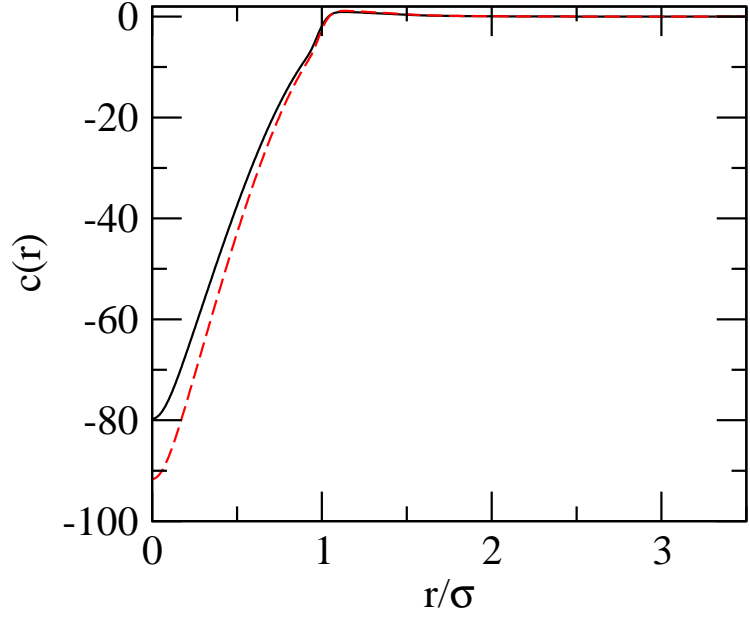}
\caption{The direct correlation function at temperature $T=1$ (solid
line) and $.8$ (dashed line) for density $\rho_0{\sigma^3}=1.1$.}
\label{fig1}
\end{figure}

\begin{figure}[htbp!]
\centering
\includegraphics[width=0.6\textwidth]{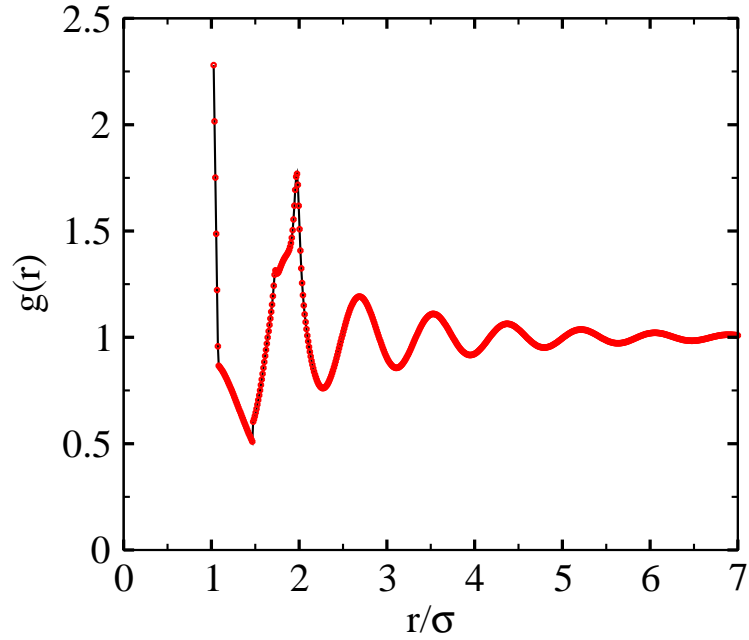}
\caption{The Bernal pair correlation function $g_B(r)$ vs.
$r/\sigma$ where $\sigma$ is the microscopic scale of the
interaction potential.} \label{fig2}
\end{figure}

For our analysis, we consider here a classical system of $N$
particles, each of mass $m$ interacting with a Lennard-Jones
potential
\be \label{LJ} u(r) = 4\epsilon \left [{\left ( \frac{\sigma}{r}
\right )}^{12}- {\left ( \frac{\sigma}{r} \right )}^{6} \right ].
\ee
The basic interaction potential in Eq. (\ref{LJ}) defines the length
scale $\sigma$ and energy scale $\epsilon$ used in defining the
units of density and temperature. The equilibrium density and the
temperature of the LJ system in the present paper will be
respectively expressed in units of $\sigma^{-3}$ and $\epsilon/k_B$.
The structure of the corresponding homogeneous liquid, denoted by
$c(r)$ is a required input in the calculation. For the LJ potential,
the direct correlation function of the uniform liquid is obtained
using the bridge function method
\cite{bridge-fn-lj,bridge-fn-lj-96}.The thermodynamic properties of
the supercooled liquid are obtained using the constant NVT ensemble
of $N$ particles interacting with the LJ potential in volume $V$ and
has a constant temperature $T$. In Fig. \ref{fig1} we show the
direct correlation function $c(r)$ obtained for density
$\rho_0\sigma^3=1.1$. The corresponding temperatures are
$k_BT/\epsilon=0.8$ and $1.0$. Next, we consider the distribution of
particle sites $\{{\bf R}_i\}$. In case of crystal, FCC lattice
serves as the particle sites. For the amorphous glassy states, the
centers for the Gaussian density profiles $\{{\bf R}_i\}$ in the
expression (\ref{dens}) for the density function are assumed to be
distributed on a random lattice. A standard procedure generally
followed \cite{ysingh,baus-colot,lowen-90} here to obtain the random
structure is to use the $g_B(R)$ corresponding to  the Bernal's
packing \cite{Bernal_1964} which is generated through the Bennett's
algorithm \cite{Bennett_algorithm}. We use the random structure
$g_s(R)$ through the following relation \cite{baus-colot}
\begin{equation}
\label{eta0} g_s(R)=g_B~[{\gamma_0}R]~~,
\end{equation}
with $\gamma_0={({\eta}/{\eta_0})}^{1/3}$ where $\eta$ denotes the
average packing fraction. $\eta_0$ is used as a scaling parameter
for the structure  such that at $\eta=\eta_0$ Bernal's structure
$g_B(R)$ is reproduced. The mapping of the function from $g_s(R)$ to
$g_B(\gamma_0{R})$ makes the structure represented by $g_s$ to
become more spread apart with increasing $\eta_0$, at a fixed
packing fraction $\eta(<\eta_0)$. The role of the $\eta_0$ on the
free energy landscape plays a crucial role in this work. We display
in Fig. \ref{fig2} the Bernal's random structure. In this regard it
should be noted that for a hard sphere system identification of the
most closely packed random structure is somewhat anomalous
\cite{Torquato_2000}. In the present context, however, the Bernal
structure is simply applied as a tool to evaluate the free energy
for an inhomogeneous density profile centered at the random set of
lattice points.

\begin{figure}[htbp!]
\centering
\includegraphics[width=0.6\textwidth]{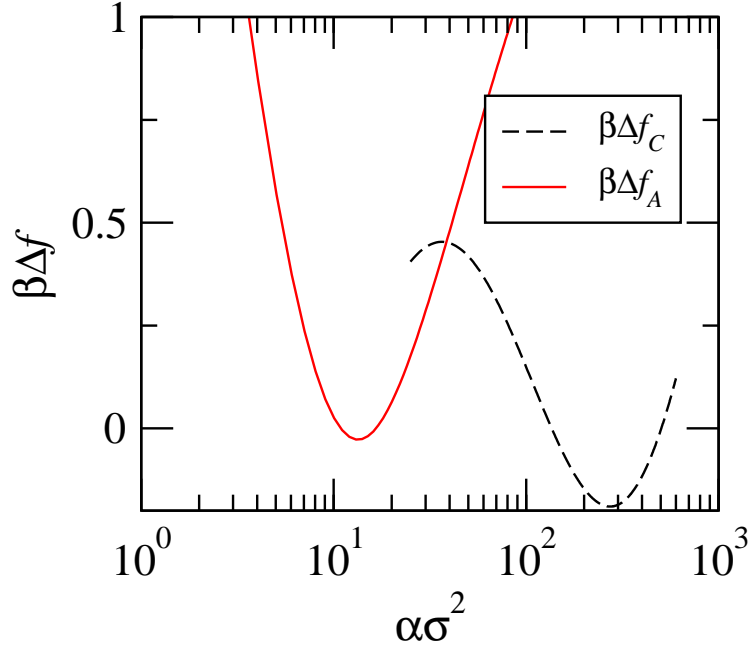}
\caption{Total free energies Vs. width parameter are shown for
density $\rho_0^{*}=1.1$. We locate the thermodynamically preferred
metastable glassy states w.r.t. the localization parameter $\alpha$
for two different structures: Bernal's random structure (solid line)
and fcc lattice sites (dashed line).} \label{fig3}
\end{figure}

\begin{figure}[htbp!]
\centering
\includegraphics[width=0.6\textwidth]{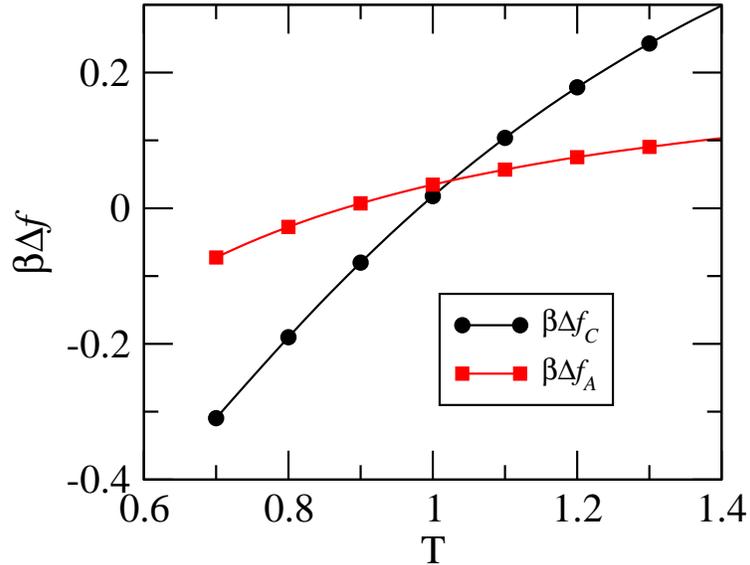}
\caption{Difference of the free energy $\Delta{f}$ of the amorphous
glassy state (solid) and fcc crystalline state (dashed) respectively
from that of the uniform liquid vs. temperature $k_BT/\epsilon$,  at
constant density $\rho_0{\sigma^3}=1.1$.} \label{fig4}
\end{figure}

\begin{figure}[htbp!]
\centering
\includegraphics[width=0.6\textwidth]{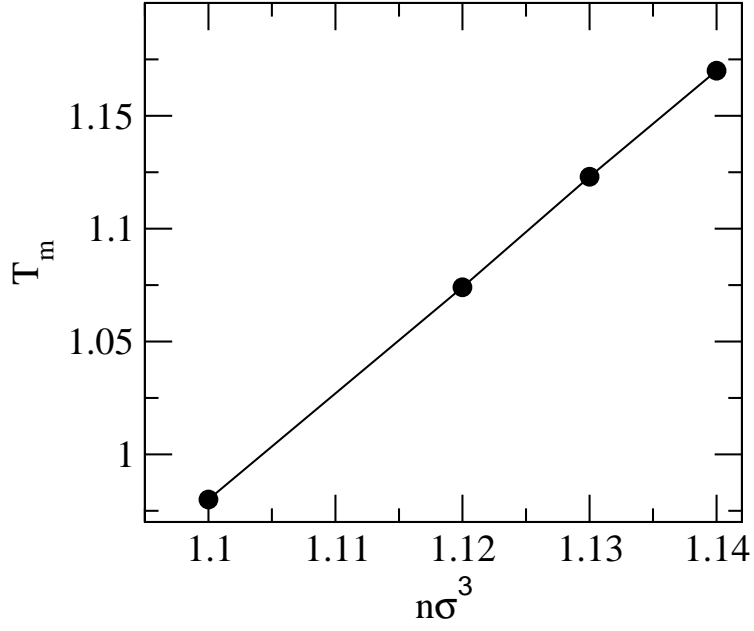}
\caption{The freezing temperature $T_m$ ( in units of
$\epsilon/k_B$) vs. density $\rho_0\sigma^3$.} \label{fig5}
\end{figure}

Using the above formulas and the input structure for the uniform
liquid in terms of $c(r)$ and the random structure $\{{\bf R}_i\}$
from the Bernal pair correlation function, the free energy is
calculated as a function of the width parameter $\alpha$. The free
energy minimum at a given temperature $T<T_m$ corresponds to a
metastable state with amorphous structure lying in the supercooled
regime. The free energy minimization with respect to $\alpha$ is
displayed in Fig. \ref{fig3} for two specific cases displaying the
crystalline and the amorphous metastable state. Note that the
metastable amorphous structure corresponds to a much lower degree of
mass localization compared to the crystalline state. The difference
of the free energy of the amorphous or the crystalline state from
that of the uniform liquid state are respectively denoted by
$\Delta{\cal F}_\mathrm{a}$ and $\Delta{\cal F}_\mathrm{c}$. The
signs of these quantities mark the relative stability of the
respective inhomogeneous state with respect to the homogeneous
liquid state. In Fig. \ref{fig4} we show that $\Delta {\cal
F}_\mathrm{c}$ become negative at temperature $T_m=0.98$ (shown with
an arrow) marking the freezing point. The amorphous state becomes
metastable compared to the liquid state at a little lower
temperature. For different choices of density $\rho_0\sigma^3$ of
the liquid, we obtain the corresponding $T_m$ as shown in Fig.
\ref{fig5}.

\subsection{Configurational Entropy}

The metastable amorphous state distinct from the uniform liquid
state, is identified by locating the intermediate minimum of the
corresponding free energy with respect to the mass localization
parameter $\alpha$. The latter determines the width of the Gaussian
density profiles in Eq. (\ref{dens}). For different temperatures,
using Eq. (\ref{gpot1-dif}) we now find the optimum free energy
differences $\Delta {\cal F}_\mathrm{a}$  and $\Delta {\cal
F}_\mathrm{v}$, respectively corresponding to the amorphous
(metastable) and the crystalline (thermodynamically stable)
structure. For the metastable states we use the Bernal's structure
to construct the random lattice $\{{\bf R}_i\}$. Different set of
lattice points are produced by varying the scaling parameter
$\eta_0$ introduced in defining the pair correlation function for
the random structure. The set of $\eta_0$ values are taken as
synonymous to different species of glass forming materials. For the
crystalline structure $\Delta {\cal F}_\mathrm{v}$ is the difference
of the free energies of the crystal and uniform liquid state. For
the crystalline state we use the fcc lattice to define the
underlying points $\{{\bf R}_i\}$. The configurational entropy
${\cal S}_c$ in the temperature range close to $T_c$ is obtained as
\begin{equation}
{\cal S}_c= {\cal S}_\mathrm{a} - {\cal S}_\mathrm{v} =
-\frac{\partial}{\partial T}{\Bigg |}_V \left [ \Delta {\cal
F}_\mathrm{a}- \Delta {\cal F}_\mathrm{v} \right ]~~.
\end{equation}
The difference between the entropies of the amorphous state with a
weak degree of mass localization ( $\alpha\sigma^2{\sim}10^1$) and
the crystalline state with sharply localized mass distribution (
$\alpha\sigma^2{\sim}10^3$) is taken here as the configurational
entropy.

\begin{figure}[htbp!]  %
\centering
\includegraphics[width=0.6\textwidth]{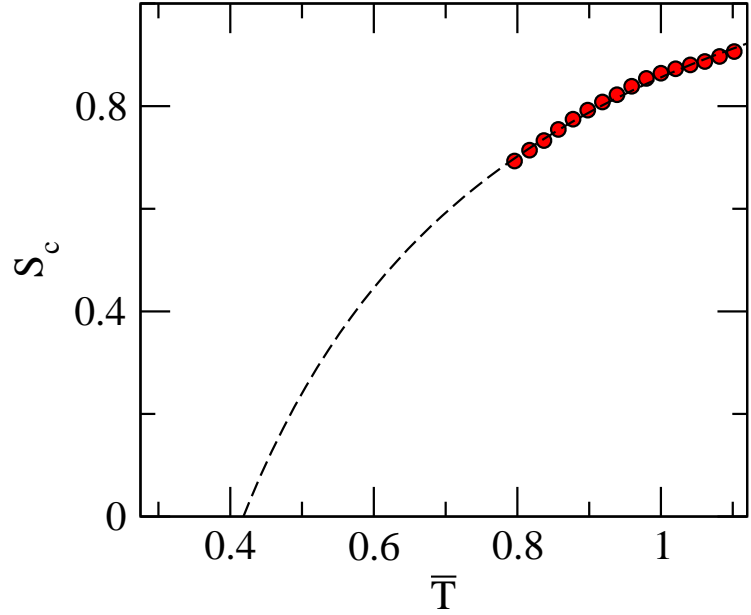}
\caption{Configurational entropy ${\cal S}_c$ Vs. $\bar{T}$
($=T/T_m$) at $\rho_0^{*}=1.1$ for amorphous structure as given by
$\eta_0=.69$. The data is extrapolated by a fit (\ref{conf-fit}) to
obtain the corresponding $T_\mathrm{K}$.} \label{fig6}
\end{figure}

\begin{figure}[htbp!]
\centering
\includegraphics[width=0.6\textwidth]{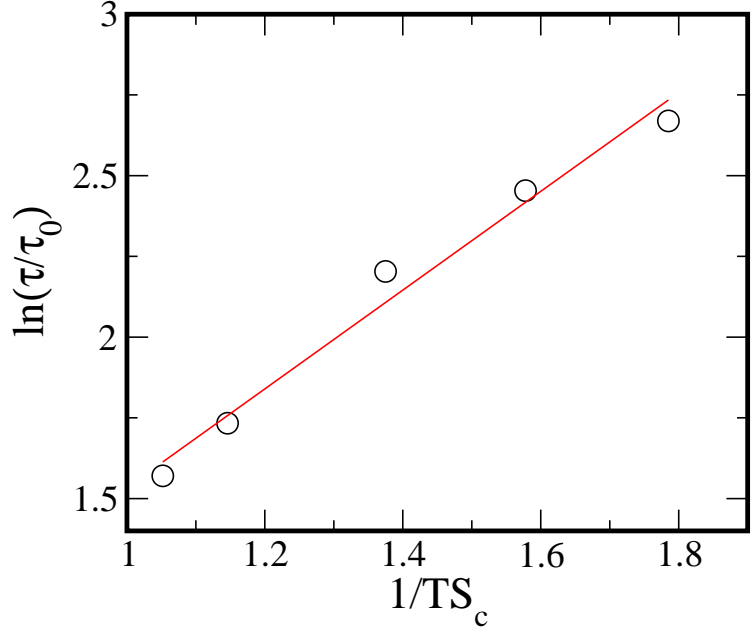}
\caption{Relaxation time $\ln[\tau/\tau_0]$ obtained from the
equations of NFH in Ref. \cite{pre_barrat} vs $T{\cal S}_c$.
Applicability of Adam-Gibbs relation given by the solid line.}
\label{fig7}
\end{figure}

In the numerical calculation, by using the free energies for the
amorphous and crystalline structures for this constant NVT ensemble,
we obtain the entropy ${\cal S}_\mathrm{a}$, of the supercooled
liquid state. At constant density $\rho_0\sigma^3=1.1$, we obtain
the ${\cal S}_c$ for a set of values for the parameter $\eta_0=.67,
.68, .69$, and $.70$. The configurational entropy studied in this
density functional model is extrapolated beyond the studied
temperature range as shown in Fig. \ref{fig6} with the form
\begin{equation}\label{conf-fit}
 {\cal S}_c={\cal S}_0 \left(1-\frac{T_\mathrm{K}}{T}\right)~~.
\end{equation}
For various $\eta_0$, we obtain by fitting the ${\cal S}_c$ to the
above form the corresponding $T_{\mathrm K}$ as well as ${\cal
S}_0$. To test the Adam-Gibbs relation we use the result for the
relaxation time $\tau$  obtained from the solution of the FNH
equations \cite{pre_barrat}. The input structure factor for the
liquid used in solving the FNH equations are same as those used in
computing the ${\cal S}_c$ in the density functional models. The
relaxation time $\tau$ is then linked to the configurational entropy
${\cal S}_c$ via AG relation so that the $\ln[\tau/\tau_0]$ vs.
$1/T{\cal S}_c$ plot is taken as the best fit to a straight line.
This is displayed in Fig. \ref{fig7}. Fitting each set of the
configurational entropy data (corresponding to a specific choice of
the parameter $\eta_0$) the Adam-Gibbs line  and hence the slope
$A_0$ for each $\eta_0$ value is obtained. Thus $A_0$, ${\cal S}_0$
and $T_\mathrm{K}$ are obtained for each $\eta_0$. Using these we
determine for the system characterized by the structure parameter
$\eta_0$, the corresponding fragility index $m$ on the Angell plot.

\subsection{Angell plot}

The plot of the glassy relaxation time $\tau$ (on a logarithmic
scale) vs. the corresponding inverse temperature $T_\mathrm{g}/T$,
(scaled with the glass transition temperature $T_\mathrm{g}$) is
referred to as the Angell plot \cite{angell,turnbull}. Here the
temperature $T_\mathrm{g}$ is defined to be the one at which the
relaxation time grows by a chosen order of magnitudes ${\cal B}$
(say) compared to its short time value for any specific system. The
quantity ${\cal B}$ is same for all materials and generally it is
chosen to be $16$ \cite{bohmer1, bohmer2}. A given curve on the
Angell plot is linked with the configurational entropy ${\cal S}_c$
of the system using the Adam-Gibbs relation.

\begin{figure}[htbp!]
\centering
\includegraphics[width=0.6\textwidth]{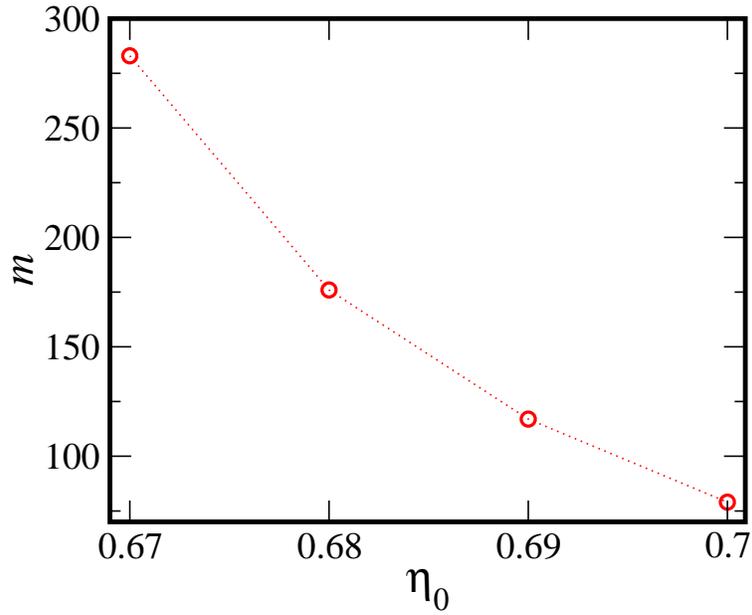}
\caption{The fragility index $m$ vs. $\eta_0$ introduced in Eq.
(\ref{eta0}).} \label{fig8}
\end{figure}

As indicated above, we have already estimated $T_\mathrm{K}$
independently from the structural data, {\em i.e.}, by an
extrapolation of the fit of the configurational entropy data
obtained at higher $T$ (near $T_m$) with the function given by
(\ref{conf-fit}). Using this form of ${\cal S}_c$ the Adam-Gibb's
relation obtains
\begin{equation}\label{AG-tau}
\tau=\tau_0 \exp\left[\frac{A_0}{{\cal S}_0\left(T-{T_\mathrm
{K}}\right)}\right].
\end{equation}
The relaxation time data is expressed as a function of the scaled
temperature $x=T_\mathrm{g}/T$ with the relation,
\be \label{aplot-data} \ln \left [ \frac{\tau}{\tau_0} \right ] =
{\cal C}_0\frac{x}{\kappa-x}~~~. \ee
We have defined the quantities ${\cal C}_0$ and $\kappa$
respectively as
\bea
\label{C0-def}
 {\cal C}_0 &=& \frac{A_0}{{{\cal S}_0}T_\mathrm{K}} \\
\label{kappa-def} \kappa &=& {T_{\mathrm g}}/{T_\mathrm{K}}~~. \eea
Using the relaxation data obtained from the solution of FNH
equations \cite{pre_barrat}, the constant ${\cal C}_0$ is
calculated. For every choice of the parameter $\eta_0$ which
characterize the structure of a particular glass forming system in
the DFT model, a corresponding $C_0$ is obtained. At
$T=T_{\mathrm{g}}$, {\em i.e.}, $x=1$ we obtain,
\be \label{beq1} {\cal B}=\ln \frac{\tau}{\tau_0}{\Big
|}_{T=T_{\mathrm{g}}} =\frac{{\cal C}_0}{\kappa-1}~~.\ee
The fragility index $m$ defined in Eq. (\ref{frag-defn}) is obtained
by calculating the derivative of the Angell curve given in Eq.
(\ref{aplot-data}) at $T=T_\mathrm{g}$, {\em i.e.}, $x=1$.
\be \label{beq2} m=\frac{\kappa {\cal C}_0}{{(\kappa -1)}^2}. \ee
In Fig. \ref{fig8} we display the fragility index $m$ vs. the
corresponding $\eta_0$ characterizing the different structures. The
figure shows that less fragile systems have higher characteristic
$\eta_0$ values, making the gaussian centers more spread out. This
represents a structure with sharply localized particles and is more
robust for the stronger liquid in which structural degradation is
hindered. To summarize the procedure, we read ${\cal S}_0$ and
$T_{\mathrm{K}}$ from the extrapolation of ${\cal S}_c$ as shown in
Fig. \ref{fig6}. $A_0$ is obtained as a fitting parameter in Fig.
\ref{fig7}. The latter involves fitting respective data of
relaxation time (from solution of FNH equations) and configurational
entropy (from DFT) with the Adam-Gibbs relation. Using these, the
constant ${\cal C}_0$ defined in Eq. (\ref{C0-def}) is obtained for
the corresponding $\eta_0$. For a chosen value of ${\cal B}$, the
relation (\ref{beq1}) determine the $\kappa$ for a corresponding
${\cal C}_0$. Each pair of $\{{\cal C}_0,\kappa\}$ is obtained for a
chosen $\eta_0$. The underlying structures $\{{\bf R}_i\}$ used in
computing the configurational entropy ${\cal S}_c$ of the
supercooled liquid correspond to chosen $\eta_0$ defined in Eq.
(\ref{eta0}).

\begin{figure}[htbp!]
\centering
\includegraphics[width=0.6\textwidth]{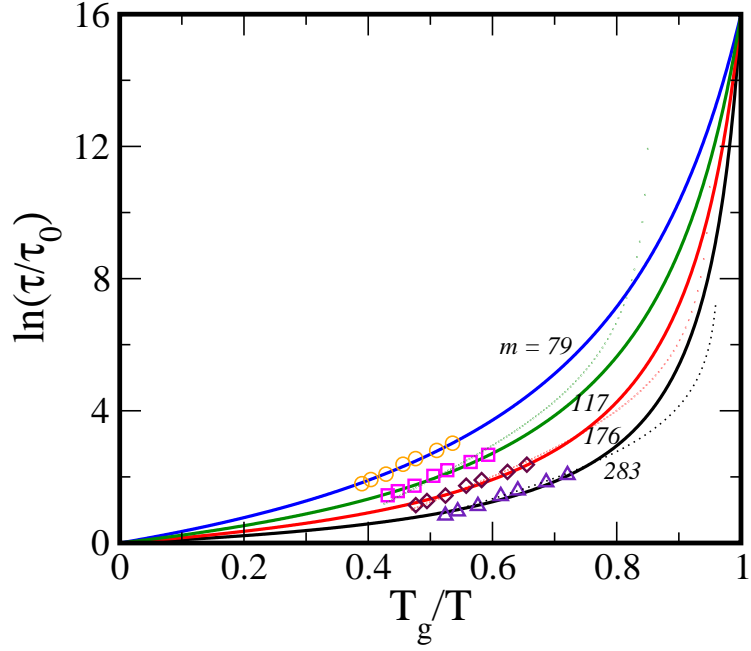}
\caption{The Angell plot of the relaxation data. Also shown in the
figure by dashed lines the corresponding power law fit predicted
from the MCT ${\sim}(T-T_c)^a$. The points shown are for $\eta_0$
=$.70$ (circles), $.69$ (squares), $.68$ (diamonds), $.67$
(triangles).}\label{fig9}
\end{figure}

\begin{figure}[htbp!]
\centering
\includegraphics[width=0.6\textwidth]{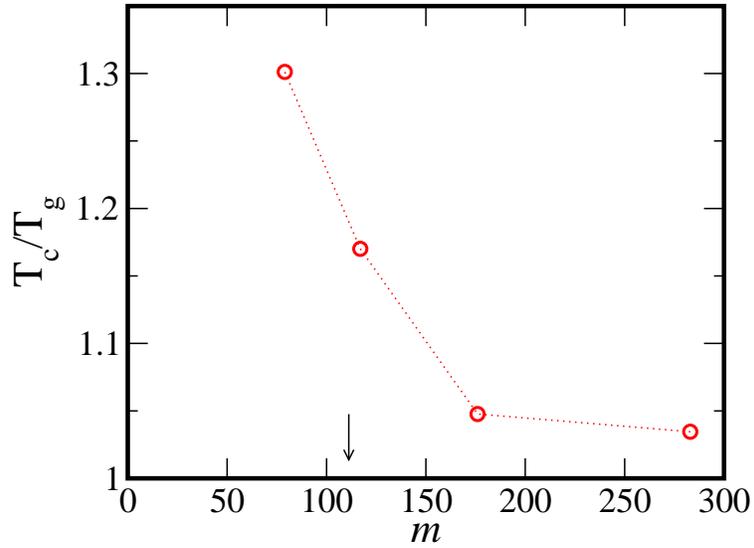}
\caption{$T_\mathrm{c}/T_\mathrm{g}$ vs. fragility index $\bar m$.
The arrow indicates the point at which the ratio is $1.3$.}
\label{fig10}
\end{figure}

Using the above result an Angell-plot of $\ln[\tau/\tau_0]$ vs. $x$
corresponding to a chosen ${\cal B}=16$ is shown in Fig. \ref{fig9}.
The different curves are characterized by respective values of
fragility $m$. The fragility is obtained using Eq. (\ref{beq2}). The
different curves correspond to a set of $\eta_0$ values. On the same
plot we display the $\tau/\tau_0$ data for the Lennard-Jones system
obtained from the solutions of the equations of FNH. The fragility
$m$ for any particular curve on the Angell plot in Fig. \ref{fig9}
is determined by the $\{\kappa,{\cal C}_0\}$. Thus we obtain $m$ for
a chosen value of $\eta_0$. The high temperature part (low values of
$x$) of each of the curves on the Angell plot in Fig. \ref{fig9}
fits well to a power law divergence $(T-T_c)^{-a}$, with a
corresponding set of $\{{T_c},a\}$. For each curve on the Angell
plot the corresponding glass transition temperature $T_g$ is
different. Since $T_\mathrm{K}$ is known, $T_\mathrm{g}$ is obtained
using Eq. (\ref{beq1}). The ratio $T_c/T_g$ vs. fragility index $m$
is shown Fig. \ref{fig10} and its inset. The agreement with
experimental results of $T_c/T_g=1.2$ \cite{my_rmp} is reached for
$m=117$ corresponding to choosing an underlying structure with
$\eta_0=.69$. This is indicated with an arrow in Fig. \ref{fig10}.

\section{Discussion}

In all its simplicity, the AG relation glues together two important
basic properties of glassy systems, respectively related to the
dynamics and the thermodynamics, making the liquid's relaxation time
to be driven by the configurational entropy. In this work, features
of the configurational entropy of the glass physics is studied
within the framework of density functional theory for the classical
liquids. Generally classical DFT has been widely used as an order
parameter model for study of the freezing of the isotropic liquid in
to a crystalline state at the freezing point $T_m$. The dynamic
behaviors exhibited by the dense fluid can be understood by studying
the equations of generalized hydrodynamics \cite{boon-yip}. The
Adam-Gibbs relation Eq. (\ref{AG-tau}) shows that as the
configurational entropy ${\cal S}_c$ becomes zero, the relaxation
time $\tau$ diverges. The outcome of the model strongly pins on the
idea that the kinetic slowdown in supercooling is a precursor of an
underlying phase transition signifying the vitrification process.
According to the Adam-Gibbs hypothesis, the relaxation of the
undercooled liquids should involve ``cooperatively rearranging
regions (CRR)''. The CRRs define the smallest size of system of
rearranging particles such that there is no smaller groups of
particles that would independently rearrange to create a new
configuration. However, with temperature the size of the CRRs
changes and is linked to an intrinsic length scale. When temperature
decreases, the motion of particles gets cooperative on a growing
length scale. The slowdown of dynamics is therefore taken to be a
collective phenomenon. From the number of possibilities of forming a
CRRs of given size the configurational entropy ${\cal S}_c$ is
obtained. By interpreting the relaxation in the deeply supercooled
state as crossing the corresponding energy barrier, the Adam-Gibbs
relation follows.

The study of the simple form of free energy functional used in DFT
shows that below freezing point $T_m$, there are inhomogeneous
states which are metastable between liquid and crystal. Extending
the ideas of the DFT, we compute the entropy ${\cal S}$ for the
inhomogeneous state. The vibrational entropy ${\cal S}_\mathrm{v}$
is obtained from the corresponding crystalline state. We obtain the
configurational entropy ${\cal S}_c$ at the supercooled temperatures
$T<T_m$ by subtracting the vibrational part from the total entropy
${\cal S}$. Since we are considering here the inhomogeneous states
corresponding to relatively low degree of mass localization
($\alpha\sigma^2<20$), keeping up to second order in the direct
functional expansion for the free energy in terms of density
fluctuations is a reasonable approximations.

The configurational entropy ${\cal S}_c$ calculated here is at
relatively higher temperatures $T$ ($<T_m$), but close to the
freezing point. $S_c$ is extrapolated to obtain the Kauzmann
temperature $T_\mathrm{K}$. For very low temperatures, close to
$T_\mathrm{g}$, the structural information for the uniform liquid is
not good enough to obtain the free energy using the simple DFT used
here. Density fluctuations are expected to be much stronger since
the deeply supercooled state is strongly heterogeneous. Extending a
low order expansion in density fluctuations for computing the free
energy at low temperatures is therefore not reliable. For hard
sphere system there are methods like MWDA \cite{denton,mwda_pre} to
consider strongly inhomogeneous states and will be considered
elsewhere.

In the present model, $\eta_0$ is a parameter used to generate the
different structures. The latter may be identified as the various
glass forming systems. The same free energy functional when tested
with random structures obtained from computer simulation studies
\cite{kim-munakata} also identified similar metastable minima with
low degree of mass localization. The various curves shown on the
Angell plot in Fig. \ref{fig9}, corresponds to ${\cal C}_0$ values
all of which are obtained by varying the structural parameter
$\eta_0$ but keeping the relaxation data same as that for the
Lennard-Jones system. This dependence can therefore be further
explored with a different sets of relaxation data for a wider
variety of glass forming materials. The variation of the $T_c$ and
the power law exponent $a$ with the fragility index $m$ obtained in
the present work is in agreement with expected non universality of
these quantities in the standard mode coupling theory. In the
present work we are able to link the structural parameter $\eta_0$
for the amorphous state to the fragility index $m$ for the
supercooled liquid.

\section*{Acknowledgement}
\label{sec_ACK} LP acknowledges CSIR, India for financial support.
SPD acknowledges support under grant 2011/37P/47/BRNS.


\newpage
\appendix
\label{app1}

\section{Evaluation of Free energy}

To perform a numerical evaluation of the Eq. (\ref{fideg}), we first
express the free energy in terms of the inhomogeneous density
profiles represented by the Gaussian of Eq. (\ref{dens}). The ideal
part free energy is now
\begin{equation}
\beta F_{id}=\int d{\bf r} \sum_{i=1}^{N}\phi_0({\bf r}-{\bf R}_i)
\left[ \ln \left( \wedge^3\sum_{j=1}^{N}\phi_0({\bf r}-{\bf
R}_j)\right) -1\right].
\end{equation}
In the free energy calculation, $\alpha$ serves as a variational
parameter and the minimization will be performed w.r.t. this
parameter. While terms involving the lattice sites $\{R_i\}$  in the
expression are taken into account through a proper counting of the
sites enclosed within corresponding shells. The number of particle
sites within a shell of radii $R$ and $R+dR$ is taken to be $4\pi
R^2 \rho_0 g_{B}(R) dR$. Thus the ideal gas free energy is
\begin{equation}
\label{f_ide} \beta F_{id}/N=\int d{\bf r} \phi_0({\bf r}) \left[
\ln \left( \wedge^3\int d{\bf R}\phi_0({\bf r}-{\bf R})
\left(\delta({\bf R})+\rho_0 g_s({\bf R})\right) \right) -1\right].
\end{equation}
Using the Gaussian form of $\phi_0$, the above Eq. (\ref{f_ide})
reduces to the following form.
\begin{eqnarray} \label{fide1}
\beta F_{id}/N &=&
\frac{3}{2}\ln\left(\wedge^2\frac{\alpha}{\pi}\right)-1 +
\left(\frac{\alpha}{\pi}\right)^{3/2} 4 \pi \int d r r^2 e^{-\alpha
r^2}\times \nonumber\\
&& \ln\left[e^{-\alpha r^2}+ \frac{\pi\gamma_0^2}{\alpha r} \int
d\bar{R}~ \bar{R}~ g_s(\bar{R})  \rho_0 \left\{e^{-\alpha(r-\gamma_0
\bar{R})^2}-e^{p-\alpha(r+ \gamma_0 \bar{R})^2}\right\}\right],
\end{eqnarray}
where $\gamma_0$ is the scaling factor defined in Eq. (\ref{eta0}).
We compute the ideal gas part free energy per particle by supplying
the Bernal's random structure in the Eq. (\ref{fide1}). In the
asymptotic limit of the large $\alpha $ case when the Gaussian
density profiles are sharply peaked around the respective lattice
sites. Assuming that there is no overlap of the Gaussian profiles
around the different sites, the ideal gas part of the free energy is
well approximated with the asymptotic formula,
\begin{equation}\label{fide-asy}
\beta F_{id}/N=-\frac{5}{2}+\frac{3}{2}\ln\left(
\wedge^2\frac{\alpha}{\pi}\right).
\end{equation}
For the excess free energy, the right hand side of the Eq.
(\ref{gpot1-dif}) as
\begin{eqnarray} \label{fint}
\beta \Delta f_{ex} &=& \rho_0 \int{r~c(r)} \left [{4\pi}r(a_0-a_1
e^{-\frac{\alpha}{2}r^2}) - a_2\int \bar{R}~g_s(\bar{R})\right.\nonumber\\
&&~~~\left.\times\left\{e^{-\frac{\alpha}{2} (r-\gamma_0
\bar{R})^2}- e^{-\frac{\alpha}{2} (r+\gamma_0 \bar{R})^2}\right
\}d\bar{R} \right]dr
\end{eqnarray}
with the constants $a_0=\frac{1}{2}$,
$a_1=\rho_0^{-1}{(\alpha/(2\pi)}^{3/2}$,
$a_2=\sqrt{2\pi\alpha}\gamma_0^2$, where
$\gamma_0={(\eta_0/\eta)}^{1/3}$. The $R$ integral is evaluated in
terms of concentric shells as in the ideal gas part.


\newpage



\end{document}